\documentclass[aps,prd,reprint,superscriptaddress,preprintnumbers,nofootinbib]{revtex4-2}

\usepackage[utf8]{inputenc}
\usepackage{amsmath,amssymb,bm}
\usepackage{booktabs}
\usepackage{graphicx}
\usepackage{xcolor}
\usepackage{hyperref}
\usepackage{microtype}
\usepackage{comment}

\graphicspath{{figures/}}

\newcommand{\dd}{\mathrm{d}}
\newcommand{\ii}{\mathrm{i}}

\begin{document}

\title{Testing the Kerr hypothesis with orbital frequencies:\\ lessons from spacelike singularities and black bounces}

\author{Hooman Neshat}
    \email{hneshat@sissa.it}
    \affiliation{SISSA, International School for Advanced Studies,
     via Bonomea 265, 34136 Trieste, Italy}
    \affiliation{IFPU, Institute for Fundamental Physics of the Universe, via Beirut 2, 34014 Trieste, Italy}
    \affiliation{INFN Sezione di Trieste,
     via Valerio 2, 34127 Trieste, Italy}
\author{Stefano Liberati}
    \email{liberati@sissa.it}
    \affiliation{SISSA, International School for Advanced Studies,
     via Bonomea 265, 34136 Trieste, Italy}
    \affiliation{IFPU, Institute for Fundamental Physics of the Universe, via Beirut 2, 34014 Trieste, Italy}
    \affiliation{INFN Sezione di Trieste,
     via Valerio 2, 34127 Trieste, Italy}

\begin{abstract}
The inner horizon of stationary black holes is subject to classical and semiclassical instabilities, suggesting that the Kerr interior should be modified. Quantum backreaction may drive the curvature to the Planck scale, producing a spacelike singularity, or halt its growth at lower curvatures, leading to a regular black bounce. We investigate the phenomenology of these two possibilities using, respectively, a G\"urses--G\"ursey geometry with a spacelike singularity and the rotating Simpson--Visser black bounce. Both geometries reproduce the light ring and ISCO frequencies of a Kerr black hole with the same ADM mass and angular momentum, by calibration in the first case and identically in the second. We find that the null and timelike convergence conditions are violated outside the horizon, up to a finite radius beyond the light ring in the G\"urses--G\"ursey geometry and everywhere in the rotating Simpson--Visser spacetime. Moreover, their fundamental scalar quasinormal modes deviate from those of Kerr, most markedly in the damping, which is enhanced in the G\"urses--G\"ursey geometry and reduced in the rotating Simpson--Visser one. Thus, agreement with Kerr in a limited set of observables is not sufficient to establish the Kerr nature of a compact object, motivating complementary observational tests.
\end{abstract}

\maketitle

\section{Introduction}

The Kerr metric is expected to describe the exterior of astrophysical rotating black holes, but there are good reasons to believe that its interior is not physically realized. The main reason is the instability of the inner horizon. In stationary geometries, like Kerr and Reissner--Nordstr\"om, the inner horizon is also a Cauchy horizon, where perturbations undergo a strong blueshift. This makes the inner horizon unstable both classically and semiclassically, so that its fate is a question that requires both classical and semiclassical analysis.

At the classical level, perturbations that remain small at the outer horizon can acquire exponentially large locally measured energy densities as they approach the inner horizon~\cite{Penrose:1969pc}. The gravitational backreaction of interacting ingoing and outgoing streams gives rise to mass inflation~\cite{Poisson:1989zz, Poisson:1990eh, Hamilton:2008zz, Carballo-Rubio:2018pmi, Carballo-Rubio:2021bpr, DiFilippo:2022qkl}, characterized by an unbounded growth of the effective internal gravitational mass and spacetime curvature. The same instability
affects geometries in which the central singularity is replaced by a nonsingular core but an inner horizon is retained~\cite{Carballo-Rubio:2021bpr}.
One proposed way to avoid the exponential growth characteristic of this instability is to construct geometries with vanishing inner horizon surface gravity while maintaining separate inner and outer horizons~\cite{CarballoRubio2022}, a strategy that has also been applied to rotating geometries~\cite{Franzin2022}.

Avoiding classical mass inflation, however, does not automatically eliminate the semiclassical inner horizon instability~\cite{McMaken2023}. A quantum state that is regular at the outer horizon, such as the Unruh state, need not remain regular at the inner horizon, where the renormalized stress-energy tensor can diverge~\cite{Zilberman2022, McMaken2023}. Semiclassical effects have also been argued to dominate over the classical ones, and studies of their backreaction have found that it tends to push the inner horizon outward, toward the outer horizon~\cite{Barcelo2021, BarceloBoyanov2022, Boyanov2026}. Quantum backreaction therefore cannot be neglected and may substantially alter the interior geometry. Semiclassical gravity can be viewed as the
leading quantum correction to the classical Einstein equations, valid as long as the curvature remains well below the Planck scale. A notable feature of the inner horizon is that, for a large black hole, the curvature there is small, while the quantum stress-energy tensor becomes large. Quantum corrections thus become important in a regime in which the
semiclassical description is still expected to apply. The subsequent evolution of the interior remains an open question, and two broad scenarios can be distinguished, depending on whether the backreaction halts the growth of the curvature while it is still far below the Planck scale.

In the first scenario, the backreaction does not halt this growth, and the semiclassical evolution leads to a spacelike singularity. McMaken and Hamilton developed a classical model of the near inner horizon geometry of an accreting rotating black hole in which mass inflation is followed by a Kasner-like collapse toward a spacelike singularity~\cite{McMakenHamilton2021}. More recently, McMaken calculated the quantum fluxes at the Kerr inner horizon and analyzed their backreaction within a framework describing mass inflation under continued accretion~\cite{McMaken2024}. This analysis suggests that the local interior geometry evolves toward a spacelike singularity, rather than toward a known regular or extremal configuration. The full semiclassical
endpoint remains uncertain, since a stress-energy tensor calculated on Kerr is no longer a controlled source once the geometry departs substantially from that background. In this scenario, the curvature eventually reaches the Planck scale, where the semiclassical description breaks down. The singularity then signals that a full theory of quantum
gravity is needed to describe this region.

In the second scenario, the backreaction halts the growth of the curvature while it is still far below the Planck scale. The lowest-order quantum corrections then remove the divergences within the semiclassical description itself, typically through stress-energy tensors that violate the classical energy conditions. A natural class of geometries realizing this possibility is that of black bounces, in which the singularity is replaced by a regular minimal surface inside the horizon~\cite{MazzaFranzinLiberati2021, Carballo-Rubio:2019nel}.

In this work, we consider one stationary rotating geometry representing each case, and compare both with Kerr. As an example of the first class, we construct a geometry within the G\"urses--G\"ursey family used in Ref.~\cite{Franzin2022}, with a radial mass function containing a pole. This produces a spacelike curvature singularity inside the outer horizon. The physical region connected to the asymptotically flat exterior terminates at this singular surface, before the formal inner root of the horizon equation is reached. As an example of the second class, we take the rotating Simpson--Visser geometry~\cite{SimpsonVisser2018, MazzaFranzinLiberati2021}, in which the singularity is replaced by a regular minimal surface controlled by a regularization parameter $L$. Both constructions model the corresponding causal structure; they do not describe its formation or constitute self-consistent solutions of the semiclassical Einstein equations.

To retain selected features of the Kerr exterior, we require both geometries to share, with a Kerr black hole of the same ADM mass~\cite{Arnowittetal1960} and angular momentum, the angular frequencies of the prograde equatorial light ring and innermost stable circular orbit (ISCO), as measured by an observer at infinity. These two orbits characterize the strong-field region of the exterior and are related to observable quantities: the ISCO frequency sets the orbital frequency at the inner edge of a thin accretion disk~\cite{PageThorne1974}, while the frequency of the prograde equatorial light ring determines the real part of the quasinormal frequencies of modes with $\ell=m$ in the limit of large $\ell$~\cite{Cardoso2009, Yang:2012he}.
Moreover, unlike the coordinate radii of these orbits, frequencies measured at infinity can be compared directly between two different geometries. In the G\"urses--G\"ursey geometry, the position of the singular surface is a free parameter, and we fix the remaining parameters of the mass function by imposing this
matching. In the rotating Simpson--Visser geometry, the two frequencies coincide with those of Kerr for every value of $L$. Each class is therefore represented by a one-parameter family of geometries. For both families, we then analyze the null and timelike convergence conditions, which govern the focusing of geodesic congruences, and compute the quasinormal modes of a massless scalar field, comparing them with Kerr.

A natural objection is that a modified interior need not have any observational consequences at all. Indeed, one can always construct a geometry that coincides exactly with Kerr outside the event horizon but differs from it inside. Within the G\"urses--G\"ursey family, it suffices to replace the mass function $m(r)$ with $\tilde m(r) = M + \chi(r)\,[m(r)-M]$, where $\chi$ is a smooth cutoff function equal to unity in the core region and vanishing identically for
all $r \geq r_c$, with $r_c$ located inside or on the Kerr outer horizon. Such a function can be infinitely differentiable, but it is necessarily non-analytic,\footnote{A real function is analytic if, in a neighborhood
of every point, it coincides with its Taylor series. On a connected domain, an analytic function is therefore determined everywhere by its values on any open interval; in particular, it cannot vanish on an interval without vanishing identically. The function equal to $e^{-1/x}$ for $x>0$ and to zero for $x\leq 0$ is the standard counterexample: it is
infinitely differentiable and all its derivatives vanish at $x=0$, so its Taylor series there is identically zero, yet the function is nonzero for every $x>0$.} and it is precisely this property that allows the deviation from Kerr to be switched off entirely at or inside the horizon. The interpolation can be chosen so that the resulting spacetime retains the interior causal structure of the model with mass function $m(r)$, while its exterior is exactly Kerr; no observation performed outside the
horizon can then distinguish it from Kerr. 
Still, it is far from clear that such composite geometries can arise from a realistic dynamical evolution. The causal shielding of the exterior relies on the outer horizon being a null, stationary boundary. Once
semiclassical effects are included, however, both horizons become dynamical~\cite{Arrechea:2026hhl}: fluxes violating the null energy condition (NEC) turn the outer horizon into a contracting, timelike surface~\cite{Arrechea:2026rua}, while at the inner horizon they may dominate over classical mass inflation~\cite{Barcelo:2022gii} and lead to an outward evolution of the latter; nothing then ensures that deviations from Kerr remain confined behind the outer horizon. In any case, we shall not consider such geometries further, since for them the main point of this paper, namely that a limited set of observations is not by itself sufficient to test the Kerr hypothesis, would hold trivially.

Both geometries considered in this work are instead described by analytic functions, as are the mass functions commonly used to construct regular black holes. A single analytic expression then describes both the interior and the exterior, and a modification of the interior necessarily extends outside the horizon. Analyticity alone, however, does not determine how large this
exterior deviation is, which depends on the specific geometry. In our two examples, the exterior is genuinely different from Kerr. For the G\"urses--G\"ursey geometries, the horizon area is larger than that of a Kerr black hole with the same mass and angular momentum, the energy per unit rest mass of the ISCO is lower, and the null and timelike convergence conditions are violated well outside the horizon, beyond the light ring. The rotating Simpson--Visser geometries are closer to Kerr: their horizon area and ISCO energy also coincide with the Kerr values, but the null convergence condition is violated everywhere. In both cases, the damping of scalar perturbations differs substantially from Kerr. A black hole whose interior has been reshaped in either of these ways can therefore carry a genuinely non-Kerr exterior and still pass tests of the Kerr hypothesis based on characteristic orbital frequencies, while other observables, such as the damping of perturbations, can distinguish it from Kerr.

The paper is organized as follows. In Sec.~\ref{sec: geometry}, we introduce the two geometries. For the G\"urses--G\"ursey geometry, we characterize the horizons and singular surface and describe the matching of the Kerr light ring and ISCO frequencies; for the rotating Simpson--Visser geometry, we show that this matching holds identically. In Sec.~\ref{sec:convergence-conditions}, we derive necessary and sufficient conditions on an arbitrary radial mass function for the null and timelike convergence conditions (NCC and TCC) to hold in the G\"urses--G\"ursey geometry, verify their static and Kerr limits, and apply them to our mass function throughout the region between the spacelike singularity and spatial infinity. We then analyze the same conditions for the rotating Simpson--Visser geometry. Finally, in Sec.~\ref{sec: qnms} we compute massless scalar quasinormal modes along both families by direct numerical
integration and compare them with the corresponding Kerr modes.

Throughout, we use geometric units $G=c=\hbar=1$ and the metric signature
$(-,+,+,+)$.

\section{Kerr-like geometries}
\label{sec: geometry}

In this section, we introduce the two geometries used in this work. The first is the G\"urses--G\"ursey Kerr-like (GG Kerr-like) geometry, in which a spacelike singularity replaces the would-be inner horizon. It represents the first scenario discussed in the Introduction. The G\"urses--G\"ursey family describes rotating geometries through a single radial mass function, and a suitable choice of this function produces the desired causal
structure in a simple closed form. The second is the rotating Simpson--Visser Kerr-like (RSV Kerr-like) geometry, which describes a black
bounce and represents the second scenario. It is one of the simplest rotating black bounces, obtained from the static Simpson--Visser metric through the Newman--Janis procedure~\cite{MazzaFranzinLiberati2021}. Both choices are made for their simplicity: they are examples of the two classes, and other geometries belonging to each class could be
considered in the same way.

\subsection{G\"urses--G\"ursey Kerr-like geometry}
\label{subsec:gg-geometry}

We consider the G\"urses--G\"ursey line element used in
Ref.~\cite{Franzin2022},
\begin{align}
\dd s^2={}&-\left(1-\frac{2m(r)r}{\Sigma}\right)\dd t^2
-\frac{4a m(r)r\sin^2\theta}{\Sigma}\,\dd t\,\dd\phi
\nonumber\\
&+\frac{\Sigma}{\Delta}\,\dd r^2+\Sigma\,\dd\theta^2
+\frac{\mathcal{A}\sin^2\theta}{\Sigma}\,\dd\phi^2,
\label{eq:metric}
\end{align}
where
\begin{align}
\Sigma&=r^2+a^2\cos^2\theta,\\
\Delta&=r^2-2m(r)r+a^2,\\
\mathcal{A}&=(r^2+a^2)^2-a^2\Delta\sin^2\theta.
\end{align}
Here $M$ is the ADM mass and $a$ is the rotation parameter, so that the ADM angular momentum is $J=aM$. For the GG Kerr-like geometry, we take the mass function
\begin{equation}
m(r)=M\left[1+\frac{\alpha}{r-s}
+\frac{\beta}{(r-s)^2}\right].
\label{eq:mass-function}
\end{equation}
Here $s$ and $\alpha$ have dimensions of length, and $\beta$ has dimensions of length squared. The mass function approaches $M$ at large radius, so the spacetime is asymptotically Kerr. At $r=s$, instead, the mass function and the Kretschmann scalar blow up.

In order to guarantee that $r=s$ is a spacelike surface, we should take a closer look at some metric functions. Near $r=s$ the term proportional to $\beta$ in the mass function dominates, so if $\beta>0$, $m(r)\to+\infty$ and $\Delta\to-\infty$ as $r\to s^+$. For the metric \eqref{eq:metric}, the causal character of a surface of constant $r$ is determined by the sign of $\Delta$. This is because its normal $n_\mu=\partial_\mu r$ has norm $g^{\mu\nu}n_\mu n_\nu=g^{rr}=\Delta/\Sigma$, where $\Sigma>0$ away from the ring $\Sigma=0$. Wherever $\Delta<0$, the normal is timelike and the surface is spacelike, while wherever $\Delta>0$, the normal is spacelike and the surface is timelike. For $\beta>0$, the surfaces of constant $r$ are therefore spacelike close to $r=s$, and the singularity is spacelike. Moreover, since $\Delta>0$ at large radius, it must vanish at least once for $r>s$, so the singular surface always lies inside a horizon. For $\beta<0$, instead, $\Delta\to+\infty$ as $r\to s^+$ and the singular surface is timelike. It would then be either naked or enclosed by both an outer and an inner horizon, and the latter would reintroduce the Cauchy horizon that this construction is meant to remove. 

We also take $\alpha<0$, so that $m(r)$ approaches $M$ from below at large radius. As shown in Sec.~\ref{sec:convergence-conditions}, this ensures that the null and timelike convergence conditions are restored at large radius, whereas for $\alpha\geq0$ they would be violated out to arbitrarily large radii.

For all configurations considered in this work, $\Delta$ has a single zero for $r>s$, which is the outer horizon $r_+$. Since $\Delta\to-\infty$ as $r\to s^-$ while $\Delta(0)=a^2>0$, $\Delta$ also has a zero at $r<s$. This second zero is only a formal root: the curvature singularity at $r=s$
separates it from the exterior and prevents access to it.

\subsubsection{Matching the Kerr light ring and ISCO}
\label{subsec:matching}

In the GG Kerr-like geometry, the parameters $\alpha$ and $\beta$ are used to make the exterior geometry phenomenologically close to Kerr. For fixed $M$, $a$, and $s$, we require that the angular frequencies of the prograde equatorial light ring and of the prograde equatorial innermost stable circular orbit (ISCO), as measured by an observer at infinity, agree with their Kerr values, for the reasons discussed in the Introduction.

Following the geodesic analysis of Ref.~\cite{Franzin2022},
equatorial radial motion is governed by
\begin{equation}
r^4 \dot{r}^{\,2}=\mathcal{R}(r),
\end{equation}
where the radial geodesic potential is
\begin{equation}
\mathcal{R}(r)
=
\left[E(r^2+a^2)-a j_z\right]^2
-\Delta(r)\left[(j_z-aE)^2+\kappa r^2\right].
\end{equation}
Here $E$ and $j_z$ denote the conserved energy and axial angular momentum of the geodesic, respectively, while $\kappa=0$ for null geodesics and $\kappa=1$ for timelike geodesics. For timelike geodesics, both conserved quantities are expressed per unit rest mass. The dot denotes differentiation with respect to an affine parameter for null geodesics and proper time for timelike geodesics. For $a>0$, prograde orbits have $j_z>0$. The angular frequency of an equatorial circular orbit measured by an
observer at infinity is $\Omega=\dd\phi/\dd t$. In terms of the conserved
quantities, it reads
\begin{equation}
\Omega=-\frac{g_{tt}\,j_z+g_{t\phi}\,E}{g_{\phi\phi}\,E+g_{t\phi}\,j_z},
\label{eq:Omega}
\end{equation}
where the metric components are evaluated at $\theta=\pi/2$. For a circular
null orbit, Eq.~\eqref{eq:Omega} reduces to $\Omega=E/j_z$.

In practice, we impose the circular null orbit conditions
\begin{equation}
\mathcal{R}=0,
\qquad
\partial_r\mathcal{R}=0,
\end{equation}
with $\kappa=0$ at the light ring radius $r_{\rm LR}$, together with
the marginally stable timelike orbit conditions
\begin{equation}
\mathcal{R}=0,
\qquad
\partial_r\mathcal{R}=0,
\qquad
\partial_r^2\mathcal{R}=0,
\end{equation}
with $\kappa=1$ at the ISCO radius $r_{\rm ISCO}$. All radial derivatives are taken at fixed $E$ and $j_z$, whose values are determined
separately for the two orbits. For $\kappa=0$, $\mathcal{R}$ is quadratic
in $E$ and $j_z$, so only the ratio $j_z/E$ is determined for the light
ring. We then require
\begin{equation}
\Omega_{\rm LR}=\Omega_{\rm LR}^{\rm Kerr},
\qquad
\Omega_{\rm ISCO}=\Omega_{\rm ISCO}^{\rm Kerr},
\label{eq:matching}
\end{equation}
where $\Omega_{\rm LR}$ and $\Omega_{\rm ISCO}$ are obtained from
Eq.~\eqref{eq:Omega} evaluated at the two orbits. To distinguish the two orbits, we denote by $E_{\rm LR}$ and $j_{\rm LR}$ the conserved quantities of the null geodesic at the light ring, and by $E_{\rm ISCO}$ and $j_{\rm ISCO}$ those of the timelike geodesic at the ISCO. The radii $r_{\rm LR}$ and $r_{\rm ISCO}$ are not fixed in advance: they are determined together with $\alpha$, $\beta$, the ratio
$j_{\rm LR}/E_{\rm LR}$, and the values of $E_{\rm ISCO}$ and
$j_{\rm ISCO}$. These seven quantities are fixed by the five orbit
conditions and the two matching conditions in Eq.~\eqref{eq:matching}.

For the representative choice
\begin{equation}
M=1,
\qquad
a=0.7,
\end{equation}
the Kerr prograde light ring and ISCO are located at
$r_{\rm LR}^{\rm Kerr}=2.0133$ and $r_{\rm ISCO}^{\rm Kerr}=3.3931$, with $\Omega_{\rm LR}^{\rm Kerr}=0.2812$ and
$\Omega_{\rm ISCO}^{\rm Kerr}=0.1439$.

Since $\alpha=\beta=0$ solves Eq.~\eqref{eq:matching} for every $s$, nontrivial solutions can branch off Kerr only where the linearization of Eq.~\eqref{eq:matching} in $(\alpha,\beta)$ becomes degenerate, which happens at $s_*\simeq0.802$. For $s<s_*$ the resulting branch has $\beta<0$ and is excluded, since the singular surface would be timelike and enclosed by an inner horizon. For $s>s_*$ it has $\beta>0$ and $\alpha<0$, and Table~\ref{tab:family} shows how it evolves: as $s$ increases, the singular surface approaches the outer horizon, with $s/r_+$ growing from about $0.49$ at $s=0.85$ to about $0.95$ at $s=2.3$, while the light ring moves outward, the ISCO moves inward, and $E_{\rm ISCO}$ decreases. This evolution stops at a turning point, $s_{\max}\simeq2.362$, where the branch folds back toward smaller $s$; along the continuation, the outer horizon, the light ring, and the ISCO approach the singular surface. We therefore restrict to $s_*<s<s_{\max}$, where the geometries form a one-parameter family labeled by $s$.

\begin{table*}[tb]
\caption{Sample of the one-parameter family of GG Kerr-like geometries for $M=1$ and $a=0.7$, obtained by imposing Eq.~\eqref{eq:matching}. For each position $s$ of the singular surface, we list the parameters $\alpha$ and $\beta$, the outer horizon $r_+$, the light ring and ISCO radii, and the ISCO energy per unit rest mass. The last row gives the corresponding Kerr values. Lengths are given in units of $M$, and $\beta$ in units of $M^2$.}
\label{tab:family}
\begin{ruledtabular}
\begin{tabular}{ccccccc}
$s$ & $\alpha$ & $\beta$ & $r_+$ & $r_{\rm LR}$ & $r_{\rm ISCO}$ & $E_{\rm ISCO}$ \\
\hline
$0.85$ & $-0.0245$ & $0.0288$ & $1.7358$ & $2.0363$ & $3.3862$ & $0.8956$ \\
$1.00$ & $-0.0931$ & $0.0990$ & $1.8033$ & $2.1061$ & $3.3620$ & $0.8925$ \\
$1.25$ & $-0.1785$ & $0.1577$ & $1.9176$ & $2.2168$ & $3.3128$ & $0.8856$ \\
$1.50$ & $-0.2270$ & $0.1616$ & $2.0363$ & $2.3219$ & $3.2493$ & $0.8758$ \\
$1.70$ & $-0.2380$ & $0.1378$ & $2.1345$ & $2.4013$ & $3.1843$ & $0.8647$ \\
$2.00$ & $-0.2036$ & $0.0782$ & $2.2846$ & $2.5064$ & $3.0491$ & $0.8375$ \\
$2.30$ & $-0.0897$ & $0.0152$ & $2.4196$ & $2.5547$ & $2.7926$ & $0.7672$ \\
\hline
Kerr & $0$ & $0$ & $1.7141$ & $2.0133$ & $3.3931$ & $0.8964$
\end{tabular}
\end{ruledtabular}
\end{table*}

In the following, we use
\begin{align}
s&=1,
&
\alpha&=-0.09305099,
\nonumber\\
\beta&=0.09896781
\label{eq:parameters}
\end{align}
solely as a representative example. The relevant radial locations are summarized in Table~\ref{tab:radial-structure}.

\begin{table}[h]
\caption{Radial structure and circular orbit quantities for Kerr and for the representative GG Kerr-like configuration of Eq.~\eqref{eq:parameters}, with $M=1$ and $a=0.7$. For Kerr, the inner root of $\Delta$ is the inner horizon; for the GG Kerr-like geometry, it is only a formal root.}
\label{tab:radial-structure}
\begin{ruledtabular}
\begin{tabular}{lcc}
Quantity & Kerr & GG Kerr-like \\
\hline
Outer horizon $r_+$ & $1.7141$ & $1.8033$ \\
Spacelike singularity $s$ & -- & $1$ \\
Inner root of $\Delta$ & $0.2859$ & $0.2092$ \\
Light ring radius $r_{\rm LR}$ & $2.0133$ & $2.1061$ \\
ISCO radius $r_{\rm ISCO}$ & $3.3931$ & $3.3620$ \\
$\Omega_{\rm LR}$ & $0.2812$ & $0.2812$ \\
$\Omega_{\rm ISCO}$ & $0.1439$ & $0.1439$ \\
$E_{\rm ISCO}$ & $0.8964$ & $0.8925$
\end{tabular}
\end{ruledtabular}
\end{table}

\begin{figure}[h]
\centering
\includegraphics[width=\columnwidth]
{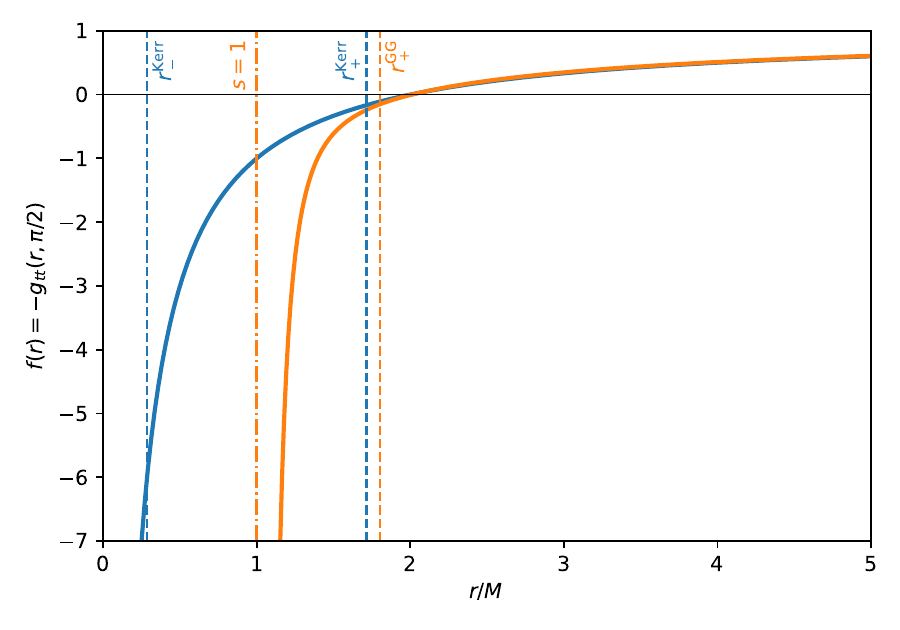}
\caption{
Equatorial metric function $f(r)$ for Kerr (blue solid) and for the
GG Kerr-like geometry of Eq.~\eqref{eq:parameters} (orange solid), with $M=1$ and $a=0.7$. Blue dashed lines mark the Kerr inner and outer horizons, the orange dashed line marks the GG Kerr-like outer horizon, and the orange dash-dotted line marks the spacelike singularity at $r=s$. The GG Kerr-like curve is shown only for $r>s$. The values of the marked radii are listed in Table~\ref{tab:radial-structure}.
}
\label{fig:metric-comparison}
\end{figure}

Figure~\ref{fig:metric-comparison} compares the equatorial
metric function $f(r)\equiv-g_{tt}(r,\pi/2)=1-2m(r)/r$
for Kerr and the GG Kerr-like geometry of Eq.~\eqref{eq:parameters}. The two functions are almost indistinguishable at large radii, and their zero crossings, which mark the ergosphere in the equatorial
plane, nearly coincide: $r\simeq2.01$ for the GG Kerr-like geometry and $r=2$ for Kerr. In the interior, instead, they differ substantially. The GG Kerr-like function diverges at $r=s$, inside its outer horizon but at a larger radius than the Kerr inner horizon, and the GG Kerr-like outer horizon lies slightly outside the Kerr one.

\subsection{Rotating Simpson--Visser Kerr-like geometry}
\label{subsec:rsv-geometry}

As an example of the second scenario, we consider the rotating
generalization of the Simpson--Visser geometry~\cite{SimpsonVisser2018}
constructed in Ref.~\cite{MazzaFranzinLiberati2021}. Introducing
\begin{equation}
\varrho=\sqrt{r^2+L^2},
\end{equation}
where $L$ is the regularization parameter, the line element can be written as
\begin{align}
\dd s^2={}&
-\left(1-\frac{2M\varrho}{\Sigma_{\mathrm{SV}}}\right)\dd t^2
-\frac{4Ma\varrho\sin^2\theta}{\Sigma_{\mathrm{SV}}}
\,\dd t\,\dd\phi
\nonumber\\
&+\frac{\Sigma_{\mathrm{SV}}}{\Delta_{\mathrm{SV}}}\dd r^2
+\Sigma_{\mathrm{SV}}\dd\theta^2
+\frac{\mathcal{A}_{\mathrm{SV}}\sin^2\theta}
{\Sigma_{\mathrm{SV}}}\dd\phi^2 ,
\label{eq:rotating-sv-metric}
\end{align}
where
\begin{align}
\Sigma_{\mathrm{SV}}
&=
\varrho^2+a^2\cos^2\theta,
\\
\Delta_{\mathrm{SV}}
&=
\varrho^2+a^2-2M\varrho,
\\
\mathcal{A}_{\mathrm{SV}}
&=
\left(\varrho^2+a^2\right)^2
-a^2\Delta_{\mathrm{SV}}\sin^2\theta.
\end{align}
For $L=0$, this is the Kerr metric. For $L>0$, one has $\varrho\geq L>0$, so that $\Sigma_{\mathrm{SV}}>0$ everywhere and the ring singularity of Kerr is removed: the surface $r=0$, where $\varrho$ reaches its minimum value $L$, is a regular minimal surface~\cite{MazzaFranzinLiberati2021}.

The horizons are located where $\Delta_{\mathrm{SV}}=0$. This happens at $\varrho=\varrho_\pm$, where $\varrho_\pm=M\pm\sqrt{M^2-a^2}$ are the same values as for the Kerr horizons. Since $\varrho\geq L$, each of these horizons exists only if $\varrho_\pm>L$, in which case it is located at $r_\pm=\sqrt{\varrho_\pm^2-L^2}$. The causal structure therefore depends on the value of $L$:
\begin{itemize}
\item For $L<\varrho_-$, both horizons exist, as in Kerr, and the geometry retains an inner horizon.
\item For $\varrho_-<L<\varrho_+$, only the outer horizon exists. Since $\Delta_{\mathrm{SV}}<0$ at $r=0$, the minimal surface is spacelike, as follows from the argument of Sec.~\ref{subsec:gg-geometry} with $g^{rr}=\Delta_{\mathrm{SV}}/\Sigma_{\mathrm{SV}}$, and lies inside the horizon. The geometry then describes a black bounce without an inner horizon.
\item For $L>\varrho_+$, there is no horizon, and the geometry no longer
describes a black hole.
\end{itemize}
The range $\varrho_-<L<\varrho_+$ therefore represents the second scenario discussed in the Introduction. For $M=1$ and $a=0.7$, this range is
$\varrho_-\simeq0.286<L<\varrho_+\simeq1.714$.

Although the metric components formally resemble those of Kerr after the replacement $r\to\varrho$, the radial differential $\dd r$ is not transformed accordingly. The rotating Simpson--Visser metric is therefore not related to Kerr by a coordinate transformation~\cite{MazzaFranzinLiberati2021}. In particular, it does not belong to the G\"urses--G\"ursey class of Eq.~\eqref{eq:metric}.

The RSV Kerr-like geometry satisfies the matching conditions of
Eq.~\eqref{eq:matching} for every value of $L$. The reason is the following. Apart from $g_{rr}$, the metric components of
Eq.~\eqref{eq:rotating-sv-metric} are those of Kerr with $r$ replaced by $\varrho$. The conditions for equatorial circular orbits involve only these components and their radial derivatives. Since $\dd\varrho/\dd r=r/\varrho$ does not vanish for $r>0$, these conditions are the same as in Kerr, written in terms of $\varrho$. The equatorial circular orbits therefore lie at the same values of $\varrho$ as in Kerr,
and have the same angular frequencies, energies, and angular momenta. In particular, the light ring and the ISCO lie outside the horizon, at $\varrho>\varrho_+>L$, so they exist for every $L<\varrho_+$. As a result, the light ring and ISCO frequencies and the ISCO energy coincide with their Kerr values. The horizon area, $4\pi(\varrho_+^2+a^2)$, also coincides with that of Kerr. The RSV Kerr-like geometries thus form a one-parameter family, labeled by $L$, in which all these quantities are exactly those of Kerr.

\section{Convergence conditions}
\label{sec:convergence-conditions}

We now investigate the convergence conditions of the two Kerr-like geometries of Sec.~\ref{sec: geometry}. We first consider the general G\"urses--G\"ursey geometry introduced in Eq.~\eqref{eq:metric}, without specifying the radial mass function $m(r)$, and then apply the results to the GG Kerr-like mass function. In contrast with energy conditions, convergence conditions are purely geometrical statements involving the Ricci tensor and therefore do not require interpreting the geometry as arising from any particular matter source. Energy conditions for rotating geometries of the G\"urses--G\"ursey class, and the structure of their Ricci tensor, have been investigated previously, including for general radial mass functions~\cite{NevesSaa2014, Torres:2016pgk, Maeda2022, Chou2025}. Here we formulate the null and timelike convergence conditions as necessary and sufficient conditions on $m(r)$ over entire constant-$r$ surfaces, and apply them to the GG Kerr-like mass function.

The null convergence condition (NCC) and timelike convergence condition (TCC) are respectively defined by
\begin{align}
R_{\mu\nu}k^\mu k^\nu &\geq 0
\qquad
\text{for every null vector } k^\mu,
\label{eq:ncc-definition}\\
R_{\mu\nu}u^\mu u^\nu &\geq 0
\qquad
\text{for every timelike vector } u^\mu.
\label{eq:tcc-definition}
\end{align}
These are the curvature conditions relevant to the focusing of null and timelike geodesic congruences, respectively \cite{BorissovaLiberatiVisser2025NCC, BorissovaLiberatiVisser2025TCC}.

For the G\"urses--G\"ursey metric, the Ricci tensor can be diagonalized in a principal orthonormal frame.\footnote{
By a principal orthonormal frame we mean an orthonormal tetrad whose basis vectors are eigenvectors of $R^\mu{}_{\nu}$. In this frame the Ricci tensor is diagonal.
}
A direct calculation gives
\begin{equation}
R_{\hat a\hat b}
=
\operatorname{diag}
\left(
\mathcal{A},
-\mathcal{A},
\mathcal{B},
\mathcal{B}
\right),
\label{eq:ricci-principal}
\end{equation}
where
\begin{align}
\mathcal{A}
&=
-\frac{r m''(r)}{\Sigma}
-\frac{2a^2\cos^2\theta\,m'(r)}{\Sigma^2},
\label{eq:ricci-A}\\
\mathcal{B}
&=
\frac{2r^2m'(r)}{\Sigma^2}.
\label{eq:ricci-B}
\end{align}
Equations~\eqref{eq:ricci-principal}--\eqref{eq:ricci-B}
allow the NCC and TCC to be analyzed for an arbitrary mass function $m(r)$.

\subsection{Null convergence condition}
\label{subsec:ncc-general}

We consider an arbitrary null vector in the principal orthonormal frame,
\begin{equation}
k^{\hat a}
=
\left(
k^{\hat 0},
k^{\hat 1},
k^{\hat 2},
k^{\hat 3}
\right).
\end{equation}
The null condition implies
\begin{equation}
\left(k^{\hat 0}\right)^2
=
\left(k^{\hat 1}\right)^2
+
\left(k^{\hat 2}\right)^2
+
\left(k^{\hat 3}\right)^2.
\label{eq:null-vector-condition}
\end{equation}
Using Eq.~\eqref{eq:ricci-principal}, the contraction entering the NCC is
therefore
\begin{align}
R_{\hat a\hat b}k^{\hat a}k^{\hat b}
&=
\mathcal{A}
\left[
\left(k^{\hat 0}\right)^2
-
\left(k^{\hat 1}\right)^2
\right]
+
\mathcal{B}
\left[
\left(k^{\hat 2}\right)^2
+
\left(k^{\hat 3}\right)^2
\right]
\nonumber\\
&=
\left(\mathcal{A}+\mathcal{B}\right)
\left[
\left(k^{\hat 2}\right)^2
+
\left(k^{\hat 3}\right)^2
\right].
\label{eq:ncc-contraction}
\end{align}
Hence, for null vectors lying in the plane spanned by the basis vectors $\hat 0$ and $\hat 1$, the contraction vanishes. A nontrivial condition arises only from null vectors with nonzero components along $\hat 2$ or $\hat 3$.

From Eqs.~\eqref{eq:ricci-A} and \eqref{eq:ricci-B},
\begin{equation}
\mathcal{A}+\mathcal{B}
=
\frac{
2\left(r^2-a^2\cos^2\theta\right)m'
-r\Sigma m''
}{
\Sigma^2
}.
\label{eq:AplusB}
\end{equation}
Since $\Sigma^2>0$ away from the ring $\Sigma=0$, the NCC is therefore
equivalent to
\begin{equation}
2\left(r^2-a^2\cos^2\theta\right)m'(r)
-r\left(r^2+a^2\cos^2\theta\right)m''(r)
\geq 0.
\label{eq:ncc-general}
\end{equation}

It is useful to obtain conditions which guarantee the NCC at every polar angle for a fixed value of $r$. Introducing
\begin{equation}
y=\cos^2\theta,
\qquad
0\leq y\leq1,
\end{equation}
the expression in Eq.~\eqref{eq:ncc-general} can be written as
\begin{equation}
r^2\left(2m'-rm''\right)
-a^2y\left(2m'+rm''\right)
\geq 0.
\label{eq:ncc-linear-y}
\end{equation}
For fixed $r$, this quantity is linear in $y$. It is therefore non-negative for every $\theta$ if and only if it is non-negative at the endpoints $y=0$ and $y=1$, corresponding respectively to the equatorial plane and the rotation axis.

On the positive-$r$ branch relevant for the black hole geometries considered here, the NCC consequently holds at every polar angle if and only if
\begin{align}
2m'-rm'' &\geq 0,
\label{eq:ncc-equator}\\
2\left(r^2-a^2\right)m'
-r\left(r^2+a^2\right)m''
&\geq 0.
\label{eq:ncc-axis}
\end{align}
The first inequality is the equatorial condition, while the second is the corresponding condition on the rotation axis. Together, Eqs.~\eqref{eq:ncc-equator} and \eqref{eq:ncc-axis} constitute necessary and sufficient conditions for the NCC to hold over an entire constant-$r$ surface.

\subsection{Timelike convergence condition}
\label{subsec:tcc-general}

We now consider the TCC. Since the sign of $R_{\mu\nu}u^\mu u^\nu$ is unchanged by a positive rescaling of $u^\mu$, we may take an arbitrary timelike vector to be normalized as
\begin{equation}
u_{\hat a}u^{\hat a}=-1.
\end{equation}
Its components therefore satisfy
\begin{equation}
\left(u^{\hat 0}\right)^2
=
1+
\left(u^{\hat 1}\right)^2
+
\left(u^{\hat 2}\right)^2
+
\left(u^{\hat 3}\right)^2.
\label{eq:timelike-vector-condition}
\end{equation}
Using again Eq.~\eqref{eq:ricci-principal}, we obtain
\begin{align}
R_{\hat a\hat b}u^{\hat a}u^{\hat b}
&=
\mathcal{A}
+
\left(\mathcal{A}+\mathcal{B}\right)
\left[
\left(u^{\hat 2}\right)^2
+
\left(u^{\hat 3}\right)^2
\right].
\label{eq:tcc-contraction}
\end{align}
For this expression to be non-negative for every timelike vector, both coefficients must be non-negative independently. Hence the TCC is equivalent to
\begin{equation}
\mathcal{A}\geq0,
\qquad
\mathcal{A}+\mathcal{B}\geq0.
\label{eq:tcc-AB}
\end{equation}
The second inequality is precisely the NCC. Thus the TCC implies the NCC, while imposing one additional independent condition.

Using Eq.~\eqref{eq:ricci-A}, the additional condition imposed by the TCC can be written as
\begin{equation}
-r\left(r^2+a^2\cos^2\theta\right)m''(r)
-2a^2\cos^2\theta\,m'(r)
\geq0 .
\label{eq:tcc-additional-general}
\end{equation}
As for the NCC, the left-hand side is linear in
$y=\cos^2\theta$. Therefore, on the positive-$r$ branch, the additional TCC condition holds for every polar angle if and only if
\begin{align}
-m'' &\geq0,
\label{eq:tcc-equator}\\
-r\left(r^2+a^2\right)m''
-2a^2m'
&\geq0.
\label{eq:tcc-axis}
\end{align}

Combining these results, the TCC holds at every polar angle on a
constant-$r$ surface if and only if
\begin{align}
2m'-rm'' &\geq0,
\label{eq:tcc-summary-1}\\
2\left(r^2-a^2\right)m'
-r\left(r^2+a^2\right)m''
&\geq0,
\label{eq:tcc-summary-2}\\
-m'' &\geq0,
\label{eq:tcc-summary-3}\\
-r\left(r^2+a^2\right)m''
-2a^2m'
&\geq0.
\label{eq:tcc-summary-4}
\end{align}
The first two inequalities are the NCC conditions, while the last two are the additional conditions imposed by the TCC.

\subsection{Static and Kerr limits}
\label{subsec:convergence-limits}

Several useful consistency checks follow immediately. In the
non-rotating limit $a\to0$, Eqs.~\eqref{eq:ncc-equator} and
\eqref{eq:ncc-axis} coincide and the NCC reduces to
\begin{equation}
2m'-rm''\geq0 .
\label{eq:ncc-spherical}
\end{equation}
Likewise, the two additional TCC conditions reduce to
\begin{equation}
-m''\geq0 .
\label{eq:tcc-spherical}
\end{equation}
Thus, in the static spherically symmetric limit, the complete
conditions are
\begin{align}
\mathrm{NCC}:&\qquad
2m'-rm''\geq0,
\\
\mathrm{TCC}:&\qquad
2m'-rm''\geq0,
\qquad
-m''\geq0,
\end{align}
in agreement with the known static results \cite{BorissovaLiberatiVisser2025NCC, BorissovaLiberatiVisser2025TCC}.

As a second consistency check, the Kerr geometry corresponds to a constant mass function,
\begin{equation}
m(r)=M,
\qquad
m'(r)=m''(r)=0.
\end{equation}
All the inequalities above are then saturated, as expected for the Kerr spacetime, which is Ricci-flat.

Equations~\eqref{eq:ncc-equator}--\eqref{eq:tcc-summary-4} therefore provide general necessary and sufficient conditions, expressed solely in terms of the first two radial derivatives of $m(r)$, for the null and timelike convergence conditions to hold throughout the G\"urses--G\"ursey class on the positive-$r$ branch.

\subsection{GG Kerr-like mass function}
\label{subsec:kerrlike-convergence}

We now specialize the general convergence conditions derived above to the GG Kerr-like mass function introduced in Eq.~\eqref{eq:mass-function}. It is convenient to define
\begin{equation}
x=r-s .
\end{equation}
The first two derivatives of the mass function are then
\begin{align}
m'(r)
&=
-M\frac{\alpha x+2\beta}{x^3}, \label{eq:kerrlike-mass-derivatives1}
\\
m''(r)
&=
2M\frac{\alpha x+3\beta}{x^4}.
\label{eq:kerrlike-mass-derivatives2}
\end{align}
We restrict in what follows to the branch $r>s$, which contains the asymptotically flat exterior region.

Substituting Eq.~\eqref{eq:kerrlike-mass-derivatives1}  and Eq.~\eqref{eq:kerrlike-mass-derivatives2}  into the general NCC condition \eqref{eq:ncc-general}, and again writing
$y=\cos^2\theta$, gives
\begin{align}
&2\left(r^2-a^2y\right)m'
-r\left(r^2+a^2y\right)m''
\nonumber\\
&\quad =
-\frac{2M}{x^4}
\Big[
\left(r^2-a^2y\right)x(\alpha x+2\beta)
\nonumber\\
&\qquad\qquad
+r\left(r^2+a^2y\right)(\alpha x+3\beta)
\Big].
\label{eq:kerrlike-ncc-contraction}
\end{align}
Since $M>0$ and $x^4>0$ away from the singular surface, the NCC at a given $(r,\theta)$ is therefore equivalent to
\begin{equation}
\hspace{-0.4cm}
-\left(r^2-a^2y\right)x(\alpha x+2\beta)
-r\left(r^2+a^2y\right)(\alpha x+3\beta)
\geq0 .
\label{eq:kerrlike-ncc-pointwise}
\end{equation}

As in the general analysis, the expression is linear in $y$.
Consequently, the NCC holds over the entire constant-$r$ surface if and only if it holds at the equator and on the rotation axis. These two conditions are
\begin{align}
-x(\alpha x+2\beta)
-r(\alpha x+3\beta)
&\geq0,
\label{eq:kerrlike-ncc-equator}\\
\hspace{-0.4cm}
-\left(r^2-a^2\right)x(\alpha x+2\beta)
-r\left(r^2+a^2\right)(\alpha x+3\beta)
&\geq0.
\label{eq:kerrlike-ncc-axis}
\end{align}

The TCC additionally requires Eq.~\eqref{eq:tcc-additional-general}. For the present mass function, its left-hand side becomes
\begin{align}
&r\left(r^2+a^2y\right)m''
+2a^2y\,m'
\nonumber\\
&\quad =
\frac{2M}{x^4}
\Big[
r\left(r^2+a^2y\right)(\alpha x+3\beta)
-a^2y\,x(\alpha x+2\beta)
\Big].
\label{eq:kerrlike-tcc-contraction}
\end{align}
The additional TCC condition is therefore equivalent to
\begin{equation}
-r\left(r^2+a^2y\right)(\alpha x+3\beta)
+a^2y\,x(\alpha x+2\beta)
\geq0.
\label{eq:kerrlike-tcc-pointwise}
\end{equation}
Since this expression is again linear in $y$, the condition holds at
every polar angle if and only if
\begin{align}
-\alpha x-3\beta
&\geq0,
\label{eq:kerrlike-tcc-equator}\\
-r\left(r^2+a^2\right)(\alpha x+3\beta)
+a^2x(\alpha x+2\beta)
&\geq0.
\label{eq:kerrlike-tcc-axis}
\end{align}
Thus, Eqs.~\eqref{eq:kerrlike-ncc-equator} and \eqref{eq:kerrlike-ncc-axis} are necessary and sufficient for the NCC to hold on a constant-$r$ surface, while the TCC additionally requires Eqs.~\eqref{eq:kerrlike-tcc-equator} and
\eqref{eq:kerrlike-tcc-axis}.

For the representative parameters in Eq.~\eqref{eq:parameters}, each of these four conditions changes sign exactly once on the branch $r>s$, at the radii
\begin{equation}
\begin{array}{c|cc}
 & \theta=\pi/2 & \theta=0 \\ \hline
\mathrm{NCC}
& 3.7410 & 3.7585 \\
\mathrm{TCC}
& 4.1908 & 4.2130
\end{array}.
\label{eq:kerrlike-convergence-roots}
\end{equation}
We denote these radii by
$r_{\mathrm{NCC}}^{(\mathrm{eq})}$,
$r_{\mathrm{NCC}}^{(\mathrm{ax})}$,
$r_{\mathrm{TCC}}^{(\mathrm{eq})}$, and
$r_{\mathrm{TCC}}^{(\mathrm{ax})}$, respectively. They are ordered as
\begin{equation}
r_+
<
r_{\mathrm{NCC}}^{(\mathrm{eq})}
<
r_{\mathrm{NCC}}^{(\mathrm{ax})}
<
r_{\mathrm{TCC}}^{(\mathrm{eq})}
<
r_{\mathrm{TCC}}^{(\mathrm{ax})},
\label{eq:kerrlike-root-ordering}
\end{equation}
where $r_+\simeq1.8033$.

The NCC is violated at every polar angle for
\begin{equation}
s<r<r_{\mathrm{NCC}}^{(\mathrm{eq})},
\end{equation}
while for
\begin{equation}
r_{\mathrm{NCC}}^{(\mathrm{eq})}
<r<
r_{\mathrm{NCC}}^{(\mathrm{ax})}
\end{equation}
it is satisfied near the equatorial plane but remains violated over a range of angles around the rotation axis. For
$r\geq r_{\mathrm{NCC}}^{(\mathrm{ax})}$, the NCC is satisfied at every polar angle.

Similarly, the TCC is violated at every polar angle for
\begin{equation}
s<r<r_{\mathrm{TCC}}^{(\mathrm{eq})},
\end{equation}
and remains violated over a finite angular range for
\begin{equation}
r_{\mathrm{TCC}}^{(\mathrm{eq})}
<r<
r_{\mathrm{TCC}}^{(\mathrm{ax})}.
\end{equation}
For $r\geq r_{\mathrm{TCC}}^{(\mathrm{ax})}$, both the NCC and TCC are satisfied over the entire constant-$r$ surface. In short, both convergence conditions are violated in a finite region that extends from the spacelike singularity to well outside the outer horizon. As Table~\ref{tab:radial-structure} shows, this region contains both the light ring, located at $r_{\rm LR}\simeq2.1061$, and the ISCO,
located at $r_{\rm ISCO}\simeq3.3620$. The two orbits whose frequencies coincide with those of Kerr therefore lie in a region where both convergence conditions are violated.

The same qualitative structure is found for all the configurations listed in Table~\ref{tab:family}: each of the four conditions changes sign exactly once for $r>s$, and the ordering~\eqref{eq:kerrlike-root-ordering} holds.
In all cases the NCC is violated beyond the light ring, and for $s\leq2$ it is also violated beyond the ISCO.

The behavior close to the singular surface and at large radius can also be understood analytically. For $x\to0^+$,
\begin{equation}
m'(r)\sim-\frac{2M\beta}{x^3},
\qquad
m''(r)\sim\frac{6M\beta}{x^4},
\end{equation}
so that, for $\beta>0$, both convergence conditions are necessarily violated sufficiently close to $r=s$. On the other hand, at large radius,
\begin{equation}
m'(r)\sim-\frac{M\alpha}{r^2},
\qquad
m''(r)\sim\frac{2M\alpha}{r^3}.
\end{equation}
For $\alpha<0$, this behavior ensures that both the NCC and TCC are restored at sufficiently large radius. Together, these two limits explain why the violations extend from the singular surface out to a finite radius, but not beyond it.

\begin{figure}[tb]
\centering
\includegraphics[width=\columnwidth]{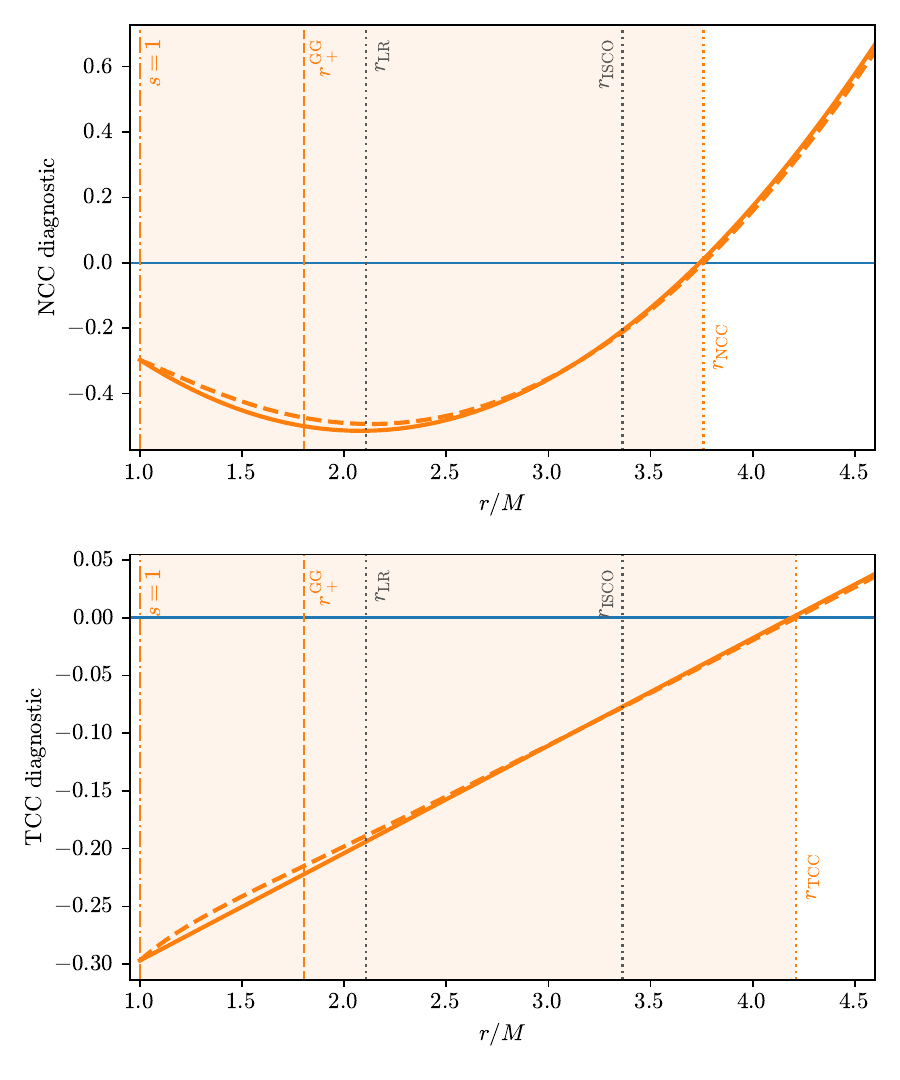}
\caption{
Quantities controlling the convergence conditions for the GG Kerr-like geometry of Eq.~\eqref{eq:parameters}, with $M=1$ and $a=0.7$. The upper panel shows the NCC conditions on the equator, Eq.~\eqref{eq:kerrlike-ncc-equator} (solid), and on the rotation axis, Eq.~\eqref{eq:kerrlike-ncc-axis} divided by $r^2+a^2$ (dashed). The lower panel shows the additional TCC
conditions on the equator, Eq.~\eqref{eq:kerrlike-tcc-equator} (solid), and on the rotation axis, Eq.~\eqref{eq:kerrlike-tcc-axis} divided by $r(r^2+a^2)$ (dashed). Dividing by these positive factors does not change
the sign of the conditions, and gives the equatorial and axial curves the same dimensions. A negative value means that the corresponding condition is violated, and a positive value means that it is satisfied. The blue horizontal line marks zero, which is the value of all these quantities for Kerr. The vertical lines mark the singular surface $r=s$ (orange
dash-dotted), the outer horizon (orange dashed), the light ring and the ISCO (grey dotted), and the axial roots $r_{\mathrm{NCC}}^{(\mathrm{ax})}$ and $r_{\mathrm{TCC}}^{(\mathrm{ax})}$ (orange dotted). The shaded region, between $r=s$ and the axial root, is where the condition is violated for at least some polar angles.
}
\label{fig:kerrlike-convergence}
\end{figure}

To illustrate these results more directly,
Fig.~\ref{fig:kerrlike-convergence} shows the radial behavior of the quantities controlling the NCC and TCC for the GG Kerr-like configuration of Eq.~\eqref{eq:parameters}. The equatorial and axial conditions are shown separately, making explicit both the region in which the convergence conditions are violated and the radii at which they are
restored. The equatorial and axial curves are very close to each other, and the corresponding roots in Eq.~\eqref{eq:kerrlike-convergence-roots} differ by less than $0.03$. The violations therefore depend only weakly
on the polar angle.

\begin{figure}[tb]
\centering
\includegraphics[width=0.8\columnwidth]{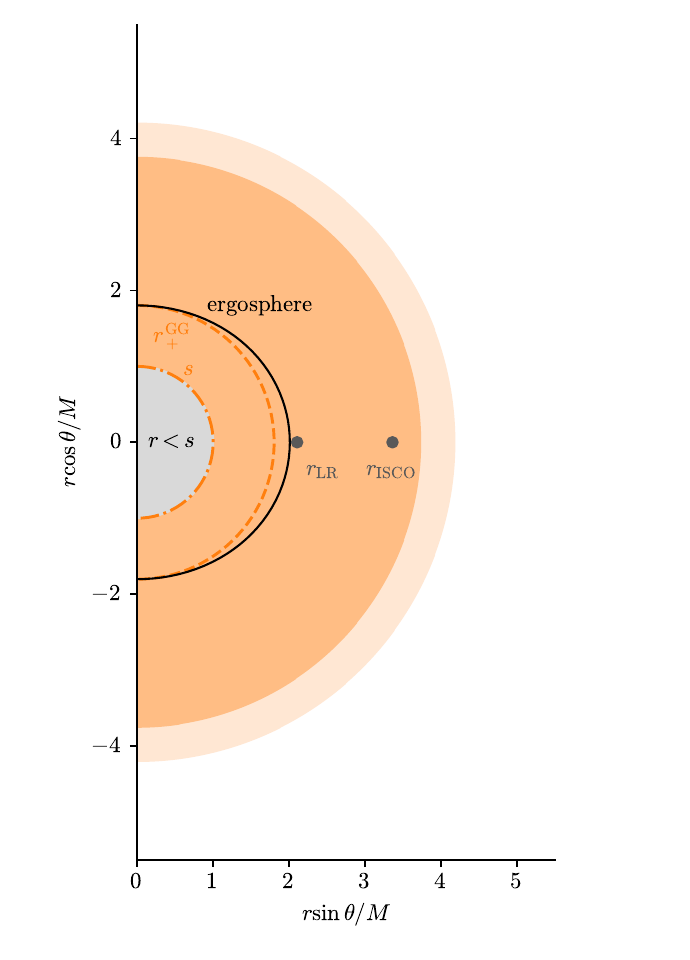}
\caption{
Regions where the convergence conditions are violated for the GG Kerr-like geometry of Eq.~\eqref{eq:parameters}, with $M=1$ and $a=0.7$, shown in a plane containing the rotation axis. Each point corresponds to a pair $(r,\theta)$ of the coordinates of Eq.~\eqref{eq:metric}: $r$ is the distance from the origin and $\theta$ the angle from the vertical axis,
which is the rotation axis. The horizontal axis lies in the equatorial plane. Dark orange marks the region where the NCC, and therefore also the TCC, is violated; light orange marks the region where the NCC is satisfied but the TCC is violated; in the white region both conditions are satisfied. The grey disk, $r<s$, is not part of the spacetime. The
orange dash-dotted and dashed curves mark the singular surface $r=s$ and the outer horizon, respectively, and the black curve marks the ergosphere. The grey dots on the equatorial plane mark the light ring and
the ISCO.
}
\label{fig:kerrlike-convergence-map}
\end{figure}

Figure~\ref{fig:kerrlike-convergence-map} summarizes these results for all polar angles. Because the equatorial and axial roots are so close, the boundaries of the violation regions are almost circles. The region where the NCC is violated surrounds the outer horizon at all polar angles
and contains the whole ergoregion, the light ring, and the ISCO. The region where only the TCC is violated forms a thin shell outside it.

\subsection{RSV Kerr-like geometry}
\label{subsec:sv-convergence}
We now turn to the RSV Kerr-like geometry of Eq.~\eqref{eq:rotating-sv-metric}. Since this geometry does not belong to the G\"urses--G\"ursey class (see Sec.~\ref{subsec:rsv-geometry}), the conditions derived in
Eqs.~\eqref{eq:ncc-equator}--\eqref{eq:tcc-summary-4}
cannot be applied to it by identifying an effective mass function.

Its convergence properties can nevertheless be determined directly. Consider the geodesic null congruence used in
Ref.~\cite{MazzaFranzinLiberati2021}, which we denote here by
$k^\mu_{\mathrm{SV}}$. The corresponding contraction with the Einstein
tensor is
\begin{equation}
G_{\mu\nu}
k^\mu_{\mathrm{SV}}k^\nu_{\mathrm{SV}}
=
-\frac{2L^2}{\Sigma_{\mathrm{SV}}^2}.
\label{eq:sv-null-contraction-einstein}
\end{equation}
Since $k^\mu_{\mathrm{SV}}$ is null,
\begin{equation}
G_{\mu\nu}
k^\mu_{\mathrm{SV}}k^\nu_{\mathrm{SV}}
=
R_{\mu\nu}
k^\mu_{\mathrm{SV}}k^\nu_{\mathrm{SV}},
\end{equation}
and hence
\begin{equation}
R_{\mu\nu}
k^\mu_{\mathrm{SV}}k^\nu_{\mathrm{SV}}
=
-\frac{2L^2}{\Sigma_{\mathrm{SV}}^2}.
\label{eq:sv-ncc-contraction}
\end{equation}
The value of this contraction depends on the normalization of
$k^\mu_{\mathrm{SV}}$, but its sign does not.
For $L>0$ one has $\Sigma_{\mathrm{SV}}>0$ everywhere, and therefore
\begin{equation}
R_{\mu\nu}
k^\mu_{\mathrm{SV}}k^\nu_{\mathrm{SV}}
<0.
\end{equation}
The existence of this null direction is sufficient to establish that the NCC is violated at every point of the RSV Kerr-like spacetime. Since the TCC implies the NCC, the TCC is also violated everywhere.

The Kerr limit is recovered for $L=0$. In that case the Ricci tensor vanishes everywhere outside the ring singularity, and both convergence conditions are saturated. Thus the RSV Kerr-like family exhibits the simple behavior
\begin{align}
L=0:\qquad&
\text{NCC and TCC are saturated},
\\
L>0:\qquad&
\text{NCC and TCC are violated everywhere}.
\label{eq:sv-convergence-summary}
\end{align}

This behavior differs from that of the GG Kerr-like geometry. In the RSV Kerr-like geometry, the violations extend over the whole spacetime for any nonzero $L$. Their magnitude, however, decreases at large radius. In particular,
Eq.~\eqref{eq:sv-ncc-contraction} behaves as
\begin{equation}
R_{\mu\nu}
k^\mu_{\mathrm{SV}}k^\nu_{\mathrm{SV}}
\sim
-\frac{2L^2}{r^4}
\qquad
(r\to\infty).
\end{equation}

\section{Scalar quasinormal modes}
\label{sec: qnms}
Perturbations provide a direct way to test how a given spacetime reacts to external disturbances. In the present work, we restrict attention to a massless scalar test field propagating on the fixed Kerr-like backgrounds of Sec.~\ref{sec: geometry} and use its quasinormal spectrum as a simple diagnostic of the departure from Kerr. Although astrophysical ringdown signals are gravitational, scalar perturbations already show how the response of the geometry depends on its structure. We compute the fundamental $(\ell,m,n)=(2,0,0)$ scalar quasinormal mode along both one-parameter families and compare it with the corresponding mode of the Kerr black hole with the same ADM mass and angular momentum, which by construction has the same light ring and ISCO frequencies, $\Omega_{\rm LR}$ and $\Omega_{\rm ISCO}$. Since the matched light ring is associated with modes with $\ell=m$, as noted in the Introduction, this $m=0$ mode is not directly controlled by it, and its quasinormal frequencies provide an independent probe of the exterior rather than a test of the matching. The wave equation governing the scalar test field reads
\begin{equation}
\Box\Phi=0.
\end{equation}
In both geometries, this equation separates as in Kerr. Using the separated ansatz
\begin{equation}
\Phi=e^{-\ii\omega t+\ii m\phi}
S_{\ell m}(\theta)R_{\ell m}(r),
\end{equation}
where $S_{\ell m}(\theta)$ denotes the angular eigenfunction and $R_{\ell m}(r)$ is the corresponding radial function, we compute the quasinormal frequencies by direct numerical integration in the frequency domain. A purely ingoing series is imposed near the outer horizon, while a
purely outgoing expansion is imposed at infinity. The two radial solutions are integrated to an intermediate radius, where their logarithmic derivatives are matched. This is equivalent to requiring that their Wronskian vanish; unlike the Wronskian itself, the mismatch does not depend on the normalization.
The complex frequency satisfying these conditions is then determined numerically.

To follow the mode from Kerr to the GG Kerr-like geometries, we introduce
\begin{equation}
m_{\lambda}(r)=M\left[1+\lambda\left(
\frac{\alpha}{r-s}+\frac{\beta}{(r-s)^2}\right)\right].
\end{equation}
At $\lambda=0$ the background is Kerr, while $\lambda=1$ gives the GG Kerr-like geometry. For $s=0.85$, close to the Kerr end of the family, we increase $\lambda$ from $0$ to $1$ in steps of $\Delta\lambda=0.01$, using the frequency found at each step as the initial seed for the next root search. We then follow the mode along the family, increasing $s$ in small steps up to $s=2.35$, again using each frequency as the seed for the next step. For the RSV Kerr-like geometry, which reduces to Kerr at $L=0$, we follow the mode in the same way by increasing $L$ in steps of $0.01$, up to $L=1.5$. We take $M=1$ and $a=0.7$. For Kerr we obtain $M\omega_{\rm Kerr}=0.50092740-0.09189685\,\ii$,\footnote{In geometric
units, $\omega$ has dimensions of inverse mass, and we quote the dimensionless combination $M\omega$. Since all parameters of the geometry scale with $M$, the frequency for a black hole of any mass follows as $\omega=(M\omega)/M$.} in agreement with the quasinormal-mode data repository of Berti and collaborators~\cite{Berti:2005ys, Berti:2009kk}
to all the digits quoted. We have also checked that our results are stable under variations of the numerical parameters of the integration.

\begin{figure}[tb]
\centering
\includegraphics[width=\columnwidth]{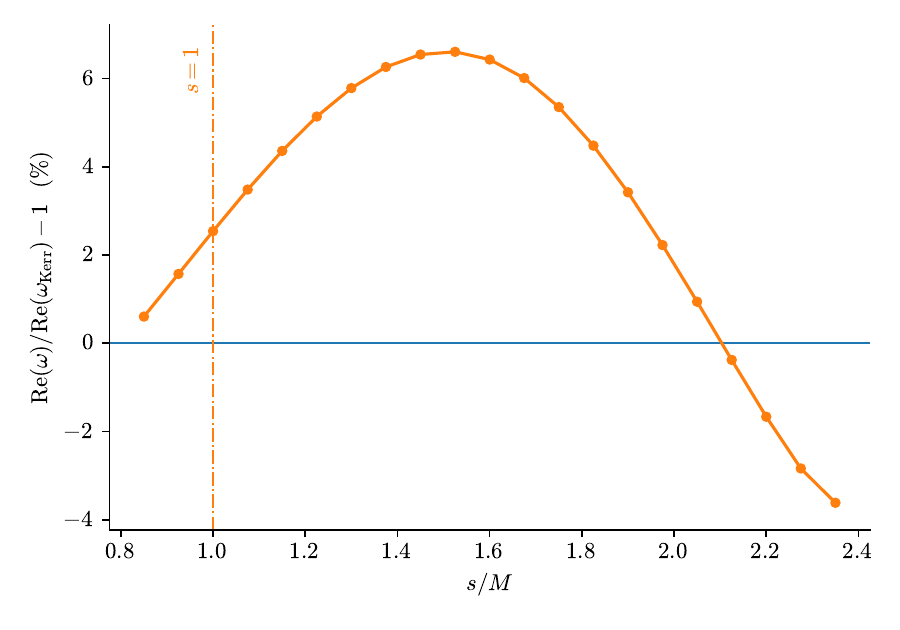}

\vspace{0.3cm}

\includegraphics[width=\columnwidth]{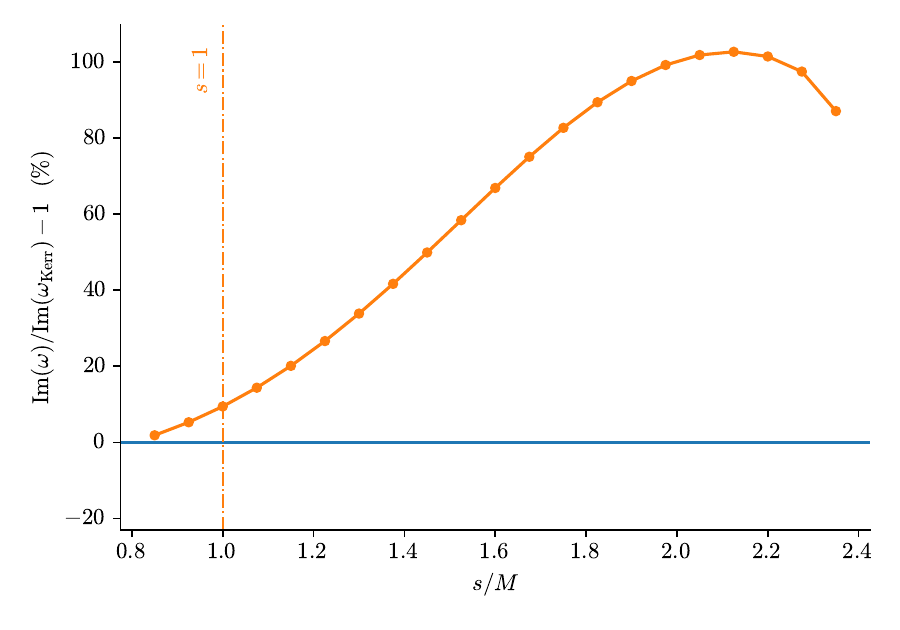}
\caption{
Relative deviation from Kerr of the real (upper panel) and imaginary (lower panel) parts of the $(\ell,m,n)=(2,0,0)$ scalar quasinormal frequency for the GG Kerr-like geometry, as a function of $s$, for $M=1$ and $a=0.7$. The blue line marks Kerr, and the orange dash-dotted line marks the representative configuration $s=1$ of Eq.~\eqref{eq:parameters}. Points are the computed values. Since $\mathrm{Im}\,\omega<0$, a positive
deviation of the imaginary part corresponds to stronger damping than in Kerr.
}
\label{fig:qnm-kerrlike}
\end{figure}

The results for the GG Kerr-like geometry are shown in
Fig.~\ref{fig:qnm-kerrlike}. The real part of the frequency first increases with $s$, reaches a maximum, and then decreases, falling below the Kerr value toward the upper end of the family. The magnitude of the imaginary part increases with $s$ over most of the family, so that these geometries are more damped than Kerr, and decreases slightly only near the
upper end. In the WKB approximation~\cite{SchutzWill1985, IyerWill1987}, the real part of a quasinormal frequency is set mainly by the height of the effective potential barrier of the radial equation, while the imaginary part is set by its curvature at the maximum. Changing $s$ modifies this barrier in a nontrivial way, since $\alpha$ and $\beta$ vary non-monotonically along the family (see Table~\ref{tab:family}). The
non-monotonic trends in Fig.~\ref{fig:qnm-kerrlike} reflect these changes.

\begin{figure}[tb]
\centering
\includegraphics[width=\columnwidth]{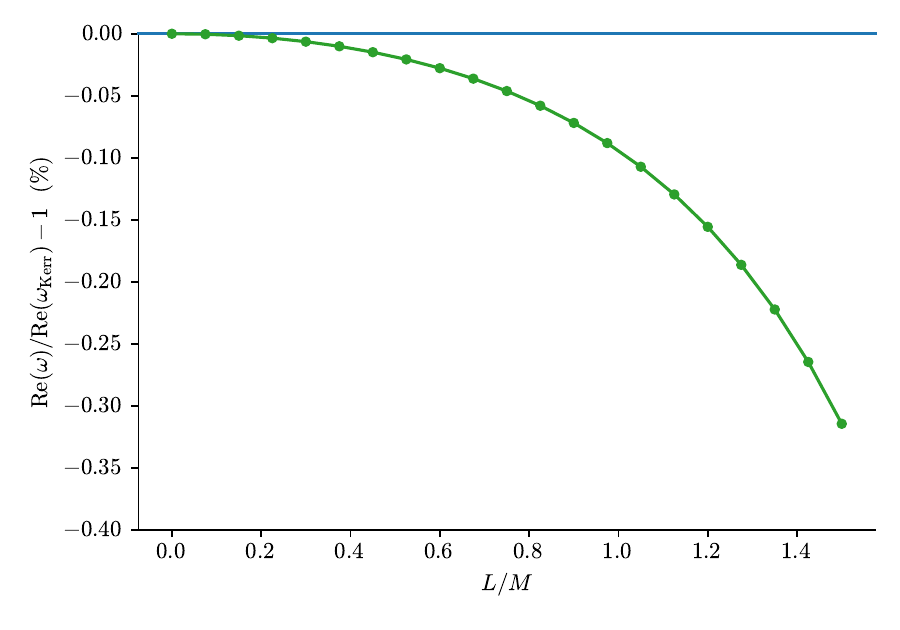}

\vspace{0.3cm}

\includegraphics[width=\columnwidth]{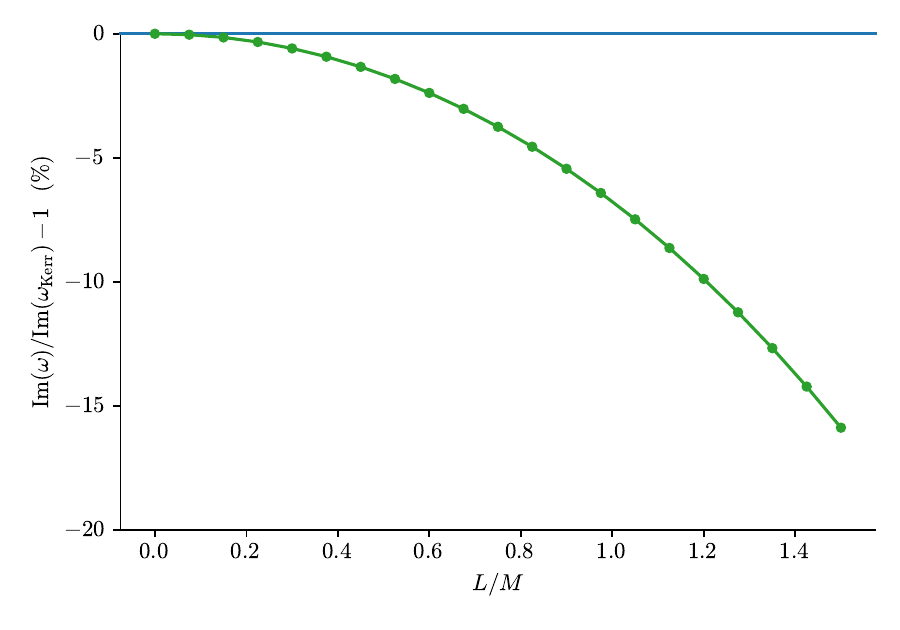}
\caption{
Relative deviation from Kerr of the real (upper panel) and imaginary (lower panel) parts of the $(\ell,m,n)=(2,0,0)$ scalar quasinormal frequency for the RSV Kerr-like geometry, as a function of the regularization parameter $L$, for $M=1$ and $a=0.7$. The blue line marks Kerr. Points are the computed values.
}
\label{fig:qnm-sv}
\end{figure}

The results for the RSV Kerr-like geometry are shown in Fig.~\ref{fig:qnm-sv}. Scalar quasinormal modes of this geometry have been studied previously~\cite{Franzin:2022iai, Khoo:2025qjc}; here we compute them along the same matched family used for the comparison with the GG Kerr-like geometry. As discussed in Sec.~\ref{subsec:rsv-geometry}, the geometries with $L<\varrho_-$ still have an inner horizon, and only those with $L>\varrho_-\simeq0.286$ are black bounces. The geometries with smaller $L$ connect the family continuously to Kerr. The real part of the frequency remains extremely close to its Kerr value, while the magnitude of the imaginary part decreases with $L$. The RSV Kerr-like geometries are therefore less damped than Kerr, whereas the GG Kerr-like geometries are more damped.

Both families reproduce the light ring and ISCO frequencies of Kerr exactly, but the real and imaginary parts of their quasinormal
frequencies respond very differently. Along the families considered, the real part deviates from its Kerr value by at most about $7\%$ for the GG Kerr-like geometries and $0.3\%$ for the RSV Kerr-like ones, whereas the damping differs by up to about $100\%$ and $16\%$, respectively, and in opposite directions. Thus, matching characteristic orbital frequencies
does not imply a scalar ringdown similar to that of Kerr, and the damping rate, $-\mathrm{Im}\,\omega$, is the quantity most sensitive to the differences between these geometries.

\section{Conclusions}

Classical and semiclassical instabilities of the inner horizon indicate that rotating black holes need not have the interior described by the Kerr solution. When quantum backreaction is included, the interior can evolve in two qualitatively different ways. If the backreaction cannot prevent
the curvature from growing to Planckian values, the evolution ends in a spacelike singularity, and a full theory of quantum gravity is needed to describe the region near it. If instead the backreaction stops this growth at sub-Planckian curvatures, the leading quantum corrections are enough to remove the divergences, and the singularity gives way to a regular black
bounce. We have treated these two possibilities on an equal footing phenomenologically, selecting one stationary rotating geometry as an example of each and comparing both with Kerr. Rather than modeling how such spacetimes form, we have used these simple examples to study how a modified interior affects the exterior.

As an example of the first scenario, we constructed the GG Kerr-like geometry, obtained within the G\"urses--G\"ursey family by introducing a radial mass function with a pole at $r=s$. The corresponding spacelike singular surface lies inside the outer horizon and terminates the physical
region connected to the asymptotically flat exterior before the formal inner root of $\Delta=0$ is reached. For fixed ADM mass, angular momentum, and singularity position, we determined the remaining parameters of the mass function by requiring the angular frequencies of the prograde equatorial light ring and ISCO, as measured by an observer at infinity, to
coincide with their Kerr values. These two frequencies were chosen because they connect the strong-field exterior to observations: the ISCO frequency marks the inner edge of thin accretion disks, while the light ring frequency determines the real part of the quasinormal frequencies of $\ell=m$ modes in the limit of large $\ell$. They are therefore natural targets for tests of the Kerr hypothesis. The matching yields a one-parameter family of geometries labeled by $s$. As an example of the second scenario, we considered the RSV Kerr-like geometry, which describes a black bounce without an inner horizon for $\varrho_-<L<\varrho_+$. For this geometry, the light ring and ISCO frequencies coincide with their Kerr values for every value of the regularization parameter $L$. Both families therefore reproduce these two Kerr frequencies exactly. Other quantities, however, behave differently in the two cases. In the GG Kerr-like geometries, the horizon area is larger than in Kerr and the ISCO energy is lower, and both
differences grow as the singular surface moves closer to the outer horizon. In the RSV Kerr-like geometries, the horizon area and the ISCO energy also coincide with their Kerr values.

We then studied the null and timelike convergence conditions. For the G\"urses--G\"ursey class, we derived necessary and sufficient conditions for these conditions to hold for an arbitrary radial mass function $m(r)$. They are expressed entirely in terms of $m'(r)$ and $m''(r)$ and reduce
to the expected results in the static and Kerr limits. Applied to the GG Kerr-like geometry, they show that both conditions are violated in a finite region that extends from the spacelike singularity to beyond the light ring and, for most of the family, beyond the ISCO. Both conditions are restored at sufficiently large radius. In the RSV Kerr-like geometry,
instead, both conditions are violated everywhere for any nonzero $L$, although the violation becomes small at large radius. In both cases, the modification of the interior is accompanied by a violation of the convergence conditions outside the horizon, in the region probed by the light ring and the ISCO.

Finally, the fundamental $(\ell,m,n)=(2,0,0)$ quasinormal mode of a massless scalar field shows a clear contrast between oscillation frequency and damping. Along both families, the oscillation frequency stays within a few percent of its Kerr value for the GG Kerr-like geometries, and within a fraction of a percent for the RSV Kerr-like ones. The damping rate, on the other hand, reaches roughly twice its Kerr value in the first case, and is reduced by about fifteen percent in the second. In these examples, it is therefore the damping, rather than the oscillation frequency, that carries the clearest signature of the departure from Kerr. These findings complement the recent observation that the first few overtones are strongly sensitive to the near-horizon geometry~\cite{Konoplya:2022pbc}, and hence, for analytic mass functions, also to the interior~\cite{DellaRocca:2026vmu}.

Overall, this body of results leads to the main message of this work. Black holes whose interior differs from Kerr, either because the inner horizon is replaced by a spacelike singularity or because the singularity is replaced by a black bounce, can reproduce the light ring and ISCO frequencies of Kerr exactly, while their exterior is genuinely different from Kerr. Such geometries would therefore pass tests of the Kerr hypothesis based on these orbital frequencies. They are nevertheless distinguishable from Kerr through other observables: in both examples considered here, the damping of scalar perturbations differs substantially from its Kerr value, and in opposite directions for the two examples. Agreement with Kerr in a limited set of observables therefore does not establish that a black hole is described by Kerr. The geometries considered here are phenomenological models, chosen for their simplicity, and not self-consistent solutions of the semiclassical Einstein equations. Other models could reproduce different Kerr properties, and thus pass different tests, while still departing from Kerr in other respects. Establishing that an astrophysical black hole is described by Kerr therefore requires combining several independent tests.

\begin{acknowledgments}
We thank Julio Arrechea and Vania Vellucci for useful discussions. 

Claude (Anthropic) was used to assist with parts of the numerical code, numerical checks, and the editing of the manuscript. The authors take full responsibility for the content of this work.
\end{acknowledgments}

\bibliography{bib-refs}

\end{document}